\documentclass[aps,prb,twocolumn,groupedaddress, showpacs]{revtex4}

\usepackage{graphicx}
\usepackage{color}

\begin{document}

\title{Determination of the magnetic domain size in the ferromagnetic
superconductor UGe$_2$ by three dimensional neutron depolarization}

\author{S. Sakarya} \email{S.Sakarya@iri.tudelft.nl}
\author{N. H. van Dijk}
\affiliation{Interfaculty Reactor Institute, Delft University of
Technology, Mekelweg 15, 2629 JB Delft, The Netherlands}

\author{E. Br\"{u}ck}
\affiliation{Van der Waals -- Zeeman Institute, University of
Amsterdam, Valckenierstraat 65, 1018 XE Amsterdam, The
Netherlands}

\date{\today}

\begin{abstract}
Three dimensional neutron depolarization measurements have been carried out on
single-crystalline UGe$_2$ between 4 K and 80 K in order to determine the average
ferromagnetic domain size~$d$. It is found that below $T_C = 52$~K uniaxial ferromagnetic
domains are formed with an estimated magnetic domain size of $d \approx 4 - 5~\mu$m.
\end{abstract}


\maketitle

\section{Introduction}

Recently, the compound UGe$_2$ has attracted much attention because superconductivity was
found to coexist with ferromagnetism. \cite{UGe2,URhGe} Until this discovery, only
superconducting compounds exhibiting antiferromagnetic order had been known like
DyMo$_6$S$_8$, GdMo$_6$S$_8$, and TbMo$_6$S$_8$. \cite{DyMo6S8,GdMo6S8,TbMo6S8}
Coexistence of antiferromagnetism and superconductivity was also found in the heavy
fermion compounds like CeIn$_3$, CePd$_2$Si$_2$, and UPd$_2$Al$_3$.
\cite{CeIn3,CePd2Si2,UPd2Al3} In these cases superconductivity and antiferromagnetism
appear simultaneously, because the Cooper pairs are insensitive to the internal fields
arising from the antiferromagnetic ordering when the superconducting coherence length
$\xi$ is much larger than the periodicity of the static antiferromagnetic ordered
structure. However, in a ferromagnetic structure we expect that the internal fields do
not cancel out on the length scale of $\xi$ and therefore have their influence on the
Cooper pairs. That ferromagnetic order excludes superconductivity, is nicely demonstrated
in ErRh$_4$B$_4$ \cite{ErRh4B4_1,ErRh4B4_2} and HoMo$_6$S$_8$ \cite{HoMo6S8}, where
standard BCS singlet-type superconductivity is suppressed when ferromagnetic order sets
in. Otherwise, if one would consider unconventional spin-triplet superconductivity
mediated by ferromagnetic spin fluctuations, the pairing is relatively insensitive to a
local magnetic field and can therefore coexist with ferromagnetic order. On the other
hand, when the ferromagnetic domain size $d$ is much smaller than the superconducting
coherence length $\xi$, one effectively has no internal magnetic field.

The coherence length $\xi$ for UGe$_2$ is estimated \cite{UGe2,Tateiwa} to be
130-200~\AA. Interestingly, Nishioka {\sl et al.}\cite{Nishioka, Nishioka2},
considering jumps in the magnetization at regular intervals of magnetic field
and at very low temperatures, estimated the ferromagnetic domain size $d$ to
be of the order of 40~\AA, by attributing the jumps to field-tuned resonant
tunnelling between quantum spin states. Since $d$ would be several times
smaller than $\xi$, it was proposed that the ferromagnetism can be cancelled
out on the scale of the coherence length of the Cooper pairs. This would imply
that the pairing mechanism for superconductivity might be of the singlet-type
after all.

UGe$_2$ crystallizes in the orthorhombic ZrGa$_2$ crystal structure (space
group Cmmm) \cite{Boulet} with unit cell dimensions $a = 4.036$~\AA, $b =
14.928$~\AA, and $c = 4.116$~\AA, containing 4 formula units. Ferromagnetic
order sets in at $T_C = 52$~K. The saturated magnetic moment at ambient
pressure is $1.4~\mu_{\mathrm{B}}$/U, directed along the $\bf a$ axis.
Magnetic measurements indicate a very strong magnetocrystalline anisotropy.
\cite{Onuki} Superconductivity is found only in a limited pressure range
between 10 and 16~kbar with a maximum transition temperature of $T_c \approx
0.7$~K. In this pressure range, the magnetic moment is still
$1~\mu_{\mathrm{B}}$/U. Within the ferromagnetic phase, a second transition
occurs at $T_X \sim 25$~K at ambient pressure, below which the magnetic moment
is enhanced. Therefore the temperature region from $T_C$ to $T_X$ was named
the weakly polarized phase, while the lower temperature region $T < T_X$ was
called the strongly polarized phase. \cite{Huxley}

In this paper we report on three dimensional neutron depolarization
measurements performed on single-crystalline UGe$_2$ at ambient pressure
between 4~K and 80~K. Our principal aim was to determine the ferromagnetic
domain size $d$ in UGe$_2$ and compare the value to the size of $\sim 40$~\AA\
estimated by Nishioka {\sl et al.}\cite{Nishioka} on the basis of the
hypothesis of field-tuned resonant tunnelling between spin quantum states.
Since the neutron is a very sensitive probe to local magnetic fields, neutron
depolarization is an excellent technique to measure the average domain size
and the domain wall width.

\section{Experiment}

\subsection{Experimental}

The measurements were performed on the Poly Axis Neutron Depolarization Analyzer (PANDA)
at the Interfaculty Reactor Institute (IRI) of the Delft University of Technology. The
used neutron wave length was 2.03~\AA\ which corresponds to a velocity of 1949 m/s.

The neutron depolarization measurements on UGe$_2$ were performed on a single-crystalline
sample with dimensions $a \times b \times c = 4.0 \times 0.440 \times 3.0 $ mm$^3$. The
$\bf b$ axis was oriented along the transmitted neutron beam ($x$) with a transmission
length $L$ and the easy axis for magnetization {\textbf a} along the vertical axis ($z$)
within the plate of the sample. The crystal has been grown from a polycrystalline ingot
using a Czochralski tri-arc technique. No subsequent heat treatment was given to the
crystal. The illuminated area was a rectangle with dimensions $y\times z = 1 \times
2$~mm$^2$ centered at the middle of the sample.

The measurements in zero field were performed during a temperature sweep from 2~K up to
80~K and down to 2~K with a low sweep rate of 10~K/hr. The measurements in non-zero field
(4 and 8~mT) were done during a similar temperature sweep with a sweep rate of 25~K/hr.
The sample was first zero-field cooled, whereafter the field was switched on at the start
of the measurements. The subsequent measurements were performed during heating and
cooling in a constant field.

\subsection{Neutron depolarization}

The neutron depolarization (ND) technique is based on the loss of polarization of a
polarized neutron beam after transmission through a (ferro)magnetic sample. Each neutron
undergoes only a series of consecutive rotations on its passage through the
(ferro)magnetic domains in the sample. It is important to note that the beam cross
section covers a huge number of domains, which results in an averaging over the magnetic
structure of the whole illuminated sample volume. This averaging causes a loss of
polarization, which is determined by the mean domain size and the mean direction cosines
of the domains. The rotation of the polarization during transmission probes the average
magnetization.

The $3 \times 3$ depolarization matrix $D$ in a ND experiment expresses the relation
between the polarization vector before ($\vec{P}^0$) and after ($\vec{P}^1$) transmission
through the sample ($\vec{P}^1 = D \vec{P}^0$). The polarization of the neutrons is
created and analyzed by magnetic multilayer polarization mirrors. In order to obtain the
complete matrix $D$, one polarization rotator is placed before the sample and another one
right after the sample. Each rotator provides the possibility to turn the polarization
vector parallel or antiparallel to the coordinate axes $x$, $y$, and $z$. The resultant
neutron intensity is finally detected by a $^3$He detector. The polarization rotators
enable us to measure any matrix element $D_{ij}$ with the aid of the intensity of the
unpolarized beam $I_S$: \begin{equation} I_S = \frac{I_{ij}+I_{-ij}}{2}. \end{equation}
where $I_{ij}$ is the intensity for $\vec{P}^0$ along $i$ and $\vec{P}^1$ along $j$. The
matrix element $D_{ij}$ is then calculated according to \begin{equation} D_{ij} =
\frac{1}{P_0} \, \frac{ I_S - I_{ij} }{I_S} \end{equation} where $P_0$ is the degree of
polarization in the absence of a sample. In our case we have $P_0 = 0.965$.

We now introduce the correlation matrix $\alpha_{ij}$: \begin{equation} \alpha_{ij} =
\Big< \int_{0}^{L} \!\!\!\! dx' \Delta B_i (x,y,z) \Delta B_j (x',y,z) \Big>,
\end{equation}where $\Delta \vec{B} (\vec{r}) = \vec{B} (\vec{r}) - \langle\vec{B}\rangle$
is the variation of the magnetic induction, $\langle \ldots \rangle$ denotes the spatial
average over the sample volume and where the integral is over the neutron transmission
length $L$ through the sample. Assuming $\alpha_{ij} \equiv 0$ for $i \neq j$ we define
the correlation function $\xi$ as\begin{equation} \xi = \sum_i \alpha_{ii}.
\end{equation} With these two quantities it can be shown that if there is no macroscopic
magnetization ($\langle \vec{B} \rangle$ = 0) the depolarization matrix is diagonal and
under the assumption of $\alpha_{ij} \equiv 0$ for $i \neq j$ given by
\cite{Rekveldt,Rosman1,Rosman2} \begin{equation} D_{ii} = e^{- \frac{\gamma^2}{v^2} L
\left\{ \xi - \alpha_{ii} \right\} } \mbox{\hspace{1cm}} i = x,y,z.
\end{equation} where $\gamma = 1.83 \times 10^8 $ s$^{-1} $T$^{-1}$ is the gyromagnetic
ratio of the neutron and $v$ its velocity.

Intrinsic anisotropy is the depolarization phenomenon that for magnetically isotropic
media the depolarization depends on the orientation of the polarization vector with
respect to the propagation direction of the neutron beam. The origin of this intrinsic
anisotropy is the demagnetization fields around magnetized volumes in the sample. In the
following we will assume that the demagnetization fields are negligible for needle shaped
magnetic domains.

We now discuss the case $\langle \vec{B} \rangle \neq 0$. When the sample shows a net
magnetization, the polarization vector will rotate in a plane perpendicular to the
magnetization direction. If the sample shape gives rise to stray fields, the rotation
angle $\phi$ is related to the net magnetization $\langle M \rangle$ by
\begin{equation} \label{phietam} \phi = \eta \frac{\gamma}{v} L \mu_0 \langle M \rangle =
\eta \frac{\gamma}{v} L \mu_0 M_S \langle m \rangle \end{equation} where $\eta$ is a
geometrically factor given in Eq.~\ref{eta} for a rectangular shaped sample and $\langle
m \rangle = M / M_S$ the reduced sample magnetization in terms of the saturation
magnetization $M_S = M_S (T)$. If the mean magnetic induction $ \langle \vec{B} \rangle$
in the sample is oriented along the $z$-axis, the depolarization matrix is, for $\phi \gg
\left( \gamma / v \right)^2 \left| \alpha_{xx} - \alpha_{yy} \right| L / 2$ (the weak
damping limit), given by \cite{Rekveldt,Rosman1,Rosman2} \begin{eqnarray}
\label{MatrixElements}  D_{xx} = D_{yy} & = & e^{- \frac{\gamma^2}{v^2} L \left\{ \xi -
\frac{\alpha_{xx}+\alpha_{yy}}{2} \right\} } \cos \phi, \nonumber \\ D_{xy} = - D_{yx} &
= & e^{- \frac{\gamma^2}{v^2} L \left\{ \xi - \frac{\alpha_{xx}+\alpha_{yy}}{2} \right\}
} \sin \phi, \nonumber \\[-0.26cm] \\[-0.26cm] D_{zz} & = & e^{- \frac{\gamma^2}{v^2} L \left\{ \xi -
\alpha_{zz} \right\} }, \nonumber
\\ D_{xz} = D_{zx} & = & D_{zy} = \ D_{yz} = 0. \nonumber  \end{eqnarray}

With the net magnetization along the $z$-axis, the rotation angle $\phi$ of the beam
polarization is obtained from the measurements by \begin{equation} \phi = \arctan \left(
\frac{D_{xy} - D_{yx}}{D_{xx} + D_{yy}} \right) \label{xigammaphi} \end{equation} and
$\xi$ is calculated with \begin{equation} \xi = - v^2 \ln \left\{ \det D \right\} / 2
\gamma^2 L. \end{equation} As mentioned earlier, ND also provides information about the
mean-square direction cosines of the magnetic induction vector in the (ferro)magnetic
domains. These are directly given by the quantities $\gamma_i = \alpha_{ii} / \xi$, where
$i = x, y, z$, and can be estimated from the measurements by
\begin{equation} \gamma_i = 1 - 2 \ln D_{ii} / \ln \left\{ \det D \right\}. \label{gamma}
\end{equation} This equation is only valid for those directions which show no net
rotation of the beam polarization.

\section{results}

\subsection{Measurements in zero field}

\begin{figure}
\includegraphics[scale=0.5]{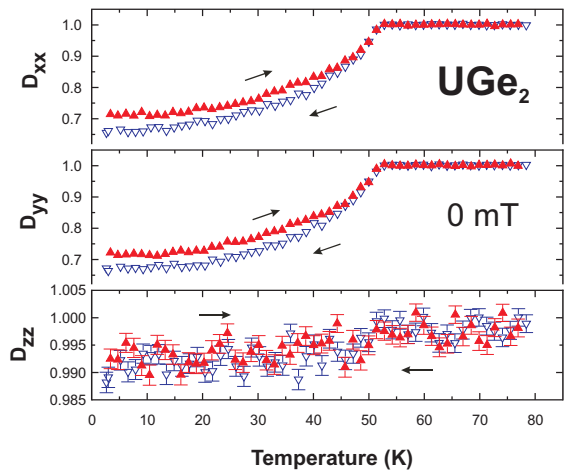}
\caption{\label{UGe2DiagElemZF} The diagonal elements of the measured depolarization
matrix $D$ for increasing and decreasing temperature for UGe$_2$. All other elements of
the depolarization matrix are zero within the experimental uncertainty. For $D_{xx}$ and
$D_{yy}$ the experimental uncertainty is within the symbol size.} \end{figure}

In Fig. \ref{UGe2DiagElemZF}  we show
the diagonal elements of the depolarization  matrix for UGe$_2$, measured in zero
magnetic field. All off-diagonal elements are zero within the experimental uncertainty in
the studied temperature range. The measurements for increasing temperature are
qualitatively the same as those for decreasing temperature, as expected.

The Curie temperature of $T_C = 52$~K is clearly indicated in Fig. \ref{UGe2DiagElemZF}
by the kink in $D_{xx}$ and $D_{yy}$. Note that $D_{xx} \equiv D_{yy}$ below $T_C$
indicates that there is no intrinsic anisotropy and hence that the magnetic domains
produce virtually no stray fields. Furthermore, $D_{zz} \approx 1$ indicates that all
moments are oriented along the $\bf a$ axis.

\subsection{Measurements in small field}

\begin{figure}
\includegraphics[scale=0.45]{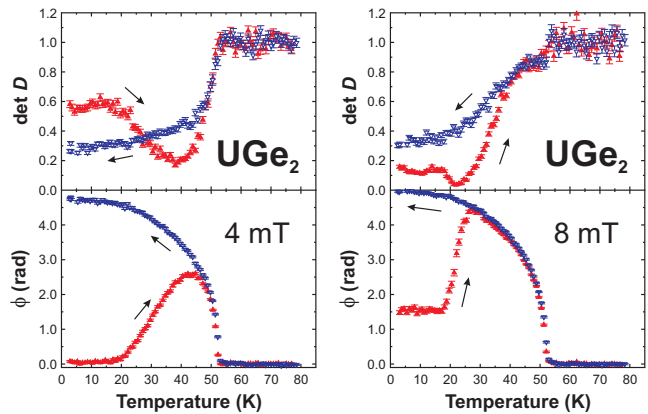}
\caption{\label{UGe2Field} The determinant of the measured depolarization matrix $\det D$
and the rotation angle $\phi$ of the beam polarization after passage through the sample
of UGe$_2$ in 4~mT and 8~mT for increasing and decreasing temperature.}
\end{figure}

In Fig. \ref{UGe2Field}  we show the determinant of the depolarization matrix $\det D$ and the
rotation angle $\phi$ after passage through the sample for measurements in fields of
respectively 4 and 8~mT (after zero-field cooling). The data of $\phi$ have been
corrected by subtracting the mean value above $T_C$, since this rotation is merely due to
the applied field.

At low temperatures the magnetic fields (applied after zero field cooling) are too small
to fully align the magnetic domains. Therefore, the measurements for increasing and
decreasing temperature do not yield the same results. Whereas for increasing temperature
the rotation shows an increase, for decreasing temperature the data represent a
monotonous magnetization curve, as expected for a field-cooled ferromagnet. Close to
$T_C$ there is no difference between field cooling or field warming.

Fig.~\ref{UGe2Field} shows that for 4~mT the depolarization is at the same level as for
0~mT. Above $T_X$, however, extra depolarization occurs. This means the system gets more
inhomogeneous, i.e. the domains grow and the magnetic correlation length increases ($\xi$
in Eq.~\ref{MatrixElements}), leading to extra depolarization. Close to $T_C$ the
depolarization disappears, because the magnetic moment decreases sharply. For decreasing
temperature the determinant has the same shape as in the case of 0~mT. At 8~mT the
determinant is already reduced below $T_X$, indicating larger domains.

Again, the Curie temperature of $T_C = 52$~K is clearly indicated by the kink
in $\det D$ and $\phi$. Also note the abrupt increase in $\phi$ around $T_X
\approx 20$~K. Evidently the system passes, with increasing temperature, from
a strongly polarized phase to a weakly polarized phase, as reported earlier.
\cite{Huxley}

\section{discussion}

\subsection{Model and results}

The measurements confirm that UGe$_2$ is a highly anisotropic uniaxial ferromagnet.
Further, the magnetic domains are long compared to their (average) width, because $D_{xx}
\equiv D_{yy}$ indicates relatively weak stray fields produced by the magnetic domains.
This allows us to assume $ \vec{B} (\vec{r}) = \mu_{0} \vec{M} (\vec{r}) $ inside the
domains. In order to analyze the data we consider a model where the sample is split into
$N$ long needles along the $\bf a$ axis with a fixed width $\Delta$ and a magnetic
induction $B_S = \mu_0 M_S$ along the $\bf a$ axis. With $N_{\uparrow}$
($N_{\downarrow}$) the number of domains with a magnetic induction pointing upwards
(downwards), we can define the reduced macroscopic magnetization of the sample, pointing
along the $z$-direction, as \begin{equation} \langle m_z \rangle = \frac{N_{\uparrow} -
N_{\downarrow}}{N_{\uparrow} + N_{\downarrow}} = \frac{\langle B_z \rangle}{ B_S}
\end{equation} Each needle will have magnetic induction $\uparrow$ or $\downarrow$ with
probability $p_\uparrow = (1 + \langle m_z \rangle)/2$ and $p_\downarrow = (1 - \langle
m_z \rangle)/2$, respectively. The polarized neutron beam traversing the sample will
therefore see a binomial distribution of $\uparrow$ and $\downarrow$, which results in a
depolarization matrix $D$ with elements \begin{eqnarray} D_{xx} = D_{yy} & = & e^{-
\frac{\gamma^2 B^2_S L}{2 v^2} \Delta \left( 1 - \langle m_z \rangle^2 \right) } \cos
\left( \frac{\gamma B_S L}{ v} \langle m_z \rangle \right), \nonumber \\ D_{xy} = -
D_{yx} & = & e^{- \frac{\gamma^2 B^2_S L}{2 v^2} \Delta \left( 1 - \langle m_z \rangle^2
\right) } \sin \left( \frac{\gamma B_S L}{ v} \langle m_z \rangle \right), \nonumber \\
D_{zz} & = & 1, \label{modelUGe2} \end{eqnarray} and all other elements equal to 0. (Note
that, since we have not taken into account the macroscopic stray fields, the angle $\phi$
should be corrected by the factor of $\eta$ (Eq.~\ref{phietam}) before calculating
$\langle m_z \rangle$ in Eq.~\ref{modelUGe2}.)

Within this binomial distribution model it is easy to show that for the case $
\langle m_z \rangle = 0 $ the average ferromagnetic domain size $d$ is equal
to $2 \Delta$. Given a domain wall (i.e. two adjacent needles with an opposite
magnetic induction), the probability of forming a domain of $n$ needles is
$\left( \frac{1}{2} \right)^n$ and the average is calculated by
$\sum_{n=1}^{\infty} n \left( \frac{1}{2} \right)^n = 2$. When a field is
applied, we have to distinguish between a domain (with size $d_{\uparrow}$) in
which the magnetic induction is parallel to the field and a domain (with size
$d_{\downarrow}$) with opposite induction. The probability of forming a domain
of size $n$ is $p_{\uparrow}^{n-1} p_{\downarrow}$ = $p_{\uparrow}^{n-1} (1 -
p_{\uparrow})$ which leads to $d_{\uparrow} / \Delta = 1 / (1 - p_{\uparrow})
= 2 / (1 - \langle m_z \rangle)$. Similarly, $d_{\downarrow} / \Delta = 2 / (1
+ \langle m_z \rangle)$.

In order to estimate the domain wall thickness $\delta$ we assume that $m_z$ changes
sinusoidally from +1 to -1 over a distance $\delta$ in the form of a Bloch wall. The
consequence is that $D_{zz}$ is slightly less than 1 in the ordered state. For such a
domain wall it is straightforward to show that the domain wall thickness $\delta$ can be
estimated by \begin{equation} \label{eqwall} \gamma_z = \langle m_z^2 \rangle = 1 -
\frac{1}{2} \left( \frac{\delta}{\Delta} \right), \end{equation} which can be measured
directly by Eq.~\ref{gamma}.

For the values of $B_S$ needed in Eq.~\ref{modelUGe2}, we use the experimental magnetic
moment of Ref. \cite{Pleiderer} which we convert to magnetic induction, remembering there
are 4 formula units per unit cell. For the value of $\eta$ in Eq.~\ref{phietam} we take
$\eta = 0.6$.

From Fig. \ref{UGe2DiagElemZF} it is clear that the data for increasing and
decreasing temperature give slightly different results for the ferromagnetic
domain size $d$ in zero magnetic field. The values found for $d = 2\Delta$ are
$ 5.1 (2)~\mu$m when cooling down slowly and $4.4 (1)~\mu$m when heating up
after fast cooling. Both values are independent of temperature. These values
indicate the domain size perpendicular to the $\bf a$~axis (along the $\bf
b$~axis). Along the $\bf a$~axis which we assume the domain size is much
larger.

The magnetic domain wall thickness $\delta$, divided by the magnetic domain size $d$, is
calculated with Eq.~\ref{eqwall} from the experimental data in Fig.~\ref{UGe2DiagElemZF},
and amounts to $\delta / d = 0.047 (23)$, independent of temperature. This gives $\delta
= 0.22~\mu$m. The size of the domain wall thickness is thus found to be only a minor
fraction of the domain size.

\begin{figure}
\includegraphics[scale=0.45]{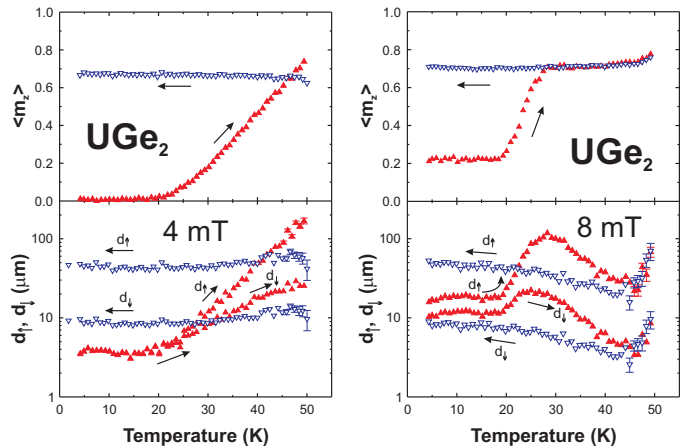}
\caption{\label{UGe2Analysis} Calculated values of the reduced macroscopic magnetization
$\langle m_z \rangle$ and the average ferromagnetic domain sizes with magnetic induction
parallel ($d_{\uparrow}$) or antiparallel ($d_{\downarrow}$) to the applied magnetic
field of 4~mT and 8~mT in UGe$_2$ for increasing and decreasing temperature.}
\end{figure}

Analysis of the data in a small magnetic field (Fig.~\ref{UGe2Field}) with
Eq.~\ref{modelUGe2} gives the results shown in Fig.~\ref{UGe2Analysis} and
table~\ref{UGe2DomainSizes}.  For 4~mT and increasing temperature (after zero field cooling), the reduced
magnetization $\langle m_z \rangle$ remains equal to 0 up to $T_X \approx 20$~K. As a
consequence $d_{\downarrow}$ is equal to $d_{\uparrow}$ and of the same order of the zero
field values. Above $T_X$ however the system gets magnetically soft and $\langle m_z
\rangle$ starts to increase linearly towards $\sim 0.7$. Domain walls are expelled above
$T_X$, since $d_{\uparrow}$ increases much faster than $d_{\downarrow}$. (Note the
logarithmic vertical scale.) While $d_{\uparrow}$ gets of the order of $100~\mu$m,
$d_{\downarrow}$ only reaches $25~\mu$m. When the domains grow in width, at a certain
moment it is no longer allowed to assume $\vec{B} (\vec{r}) = \mu_0 \vec{M} (\vec{r})$,
because stray fields produced by the domains have to be taken into account. The model
therefore is no longer appropriate close to $T_C$. \begin{table}
\caption{\label{UGe2DomainSizes}Ferromagnetic domain sizes in UGe$_2$ for increasing
temperature after zero field cooling (ZFC) and decreasing temperature in field (FC). The
sizes of the domains with magnetization parallel to the applied magnetic field is denoted
by $d_{\uparrow}$ and the domains with antiparallel magnetization by $d_{\downarrow}$.
Below $T_X$ the domain sizes are temperature independent. Above $T_X$ the domains grow.
The values shown are at a few Kelvin below $T_C$.} \begin{ruledtabular}
\begin{tabular}{c|c||c|c||c|c}
$\mu_0 H$ & Temp. & $d_{\uparrow}$~($\mu$m) & $d_{\downarrow}$~($\mu$m) & $d_{\uparrow}$~($\mu$m) & $d_{\downarrow}$~($\mu$m) \\
(mT) & incr./decr. & $T < T_X$ &  $T < T_X$  &  $T \approx T_C$  & $T \approx T_C$ \\
\hline\hline
0 & ZFC, incr. & 4.4(1) & 4.4(1) & 4.4(1) & 4.4(1) \\
0 & FC, decr.  & 5.1(2) & 5.1(2) & 5.1(2) & 5.1(2) \\
\hline
4 & ZFC, incr. & 3.9(1)  & 3.8(1) & 150(20) & 25(5) \\
4 & FC, decr.  & 46.4(8) & 9.5(2) & 60(10)  & 13(2) \\
\hline
8 & ZFC, incr. & 17.9(2) & 11.4(1) & 85(20) & 10(2)   \\
8 & FC, decr.  & 45(5)   & 8.2(1)  & 85(20) & 10(5)  \\
\end{tabular}
\end{ruledtabular}
\end{table}

For field cooling in 4~mT, the system has $\langle m_z \rangle = 0.668(1)$ for the whole
temperature range below $T_C$. The values of the domain size are shown in
table~\ref{UGe2DomainSizes}.

When after zero field cooling a field of 8~mT is turned on, the sample does get a
macroscopic magnetization, in contrast to the case of 4~mT. Up to $T_X \approx 20$~K the
reduced magnetization $\langle m_z \rangle = 0.221(2)$ is independent of temperature.
Then $\langle m_z \rangle$ starts to increase up to 0.718(3) around 30~K and is constant
afterwards up to $T_C$. When cooling down in 8~mT, $\langle m_z \rangle = 0.708(1)$ over
the whole temperature range below $T_C$.

For field warming after zero field cooling, the calculation of the domain sizes yields
unexpected temperature dependencies of domain sizes above $T_X$. As can be seen in
Fig.~\ref{UGe2Analysis}, according to the model the domain sizes grow above $T_X$ to
decrease in size again at higher temperature. Clearly there is another source of
depolarization, not accounted for by the model. Since the field is strong enough to
penetrate the sample, additional depolarization arises from an inhomogeneous magnetic
domain structure.

\subsection{Discussion}

If the domain width becomes relatively large compared to its length, stray fields become
important and the simple model assuming $\vec{B} (\vec{r}) = \mu_0 \vec{M} (\vec{r})$ is
no longer valid. Calculation of the mean-square direction cosine along the $z$ direction,
$\gamma_z$, with Eq.~\ref{gamma}, indeed shows a decrease from unity above $T_X$,
indicating that the magnetic induction $\vec{B}$ is not along the $\bf a$ axis throughout
the sample. The model can of course be improved if we no longer assume a length/width
ratio of infinity (no stray fields) for the domains. The simple model together with our
measurements, however, do show that the magnetic domain sizes in zero field are a few
$\mu$m and that by applying small fields the domains grow. Our measurements therefore
indicate that the domain sizes in UGe$_2$ at ambient pressure and down to 2~K are
certainly larger than the 40~\AA\ predicted by Nishioka {\sl et al.}\cite{Nishioka,
Nishioka2}

In Fig.~\ref{UGe2DiagElemZF} it is shown that $D_{zz}$ is less than unity below $T_C$.
This can be caused by the domain walls, but can also be accounted for by a misalignment.
A simple calculation shows that a misalignment of $8^\circ$ would fully account for the
values of $D_{zz}$ below $T_C$. The stated value of $\delta = 0.22~\mu$m (or $\delta / d
= 0.047 (23)$) should therefore be regarded as an upper limit.

From the above considerations we conclude that the domain structure of UGe$_2$ behaves
like in a conventional ferromagnet. The magnetic domain size largely exceeds the
superconducting correlation length of the Cooper pair. The magnetic domain boundaries can
therefore only give secondary effects on the superconducting order.

\section{Conclusion}

The ferromagnetic domain sizes of UGe$_2$ was studied by means of three dimensional
neutron depolarization at ambient pressure. We conclude that the existence of field-tuned
resonant tunneling between spin quantum states \cite{Nishioka,Nishioka2} is highly
unlikely. The requirement of this model is a ferromagnetic domain size of 40 \AA\, while
our measurements indicate a size a factor of 1000 larger. The observed jumps in the
magnetization should be attributed to a Barkhausen effect as discussed by Lhotel {\sl et
al.} \cite{Lhotel}. The superconductivity therefore exists within a single ferromagnetic
domain. The domain walls are not expected to strongly affect the bulk Cooper pair wave
function, as suggested by Nishioka {\sl et al.} \cite{Nishioka,Nishioka2}, since the
domain wall is less than a few percent of the average domain size.

\appendix

\section{Effect of stray fields induced by a homogeneously magnetized sample}

\begin{figure}
\includegraphics[scale=0.6]{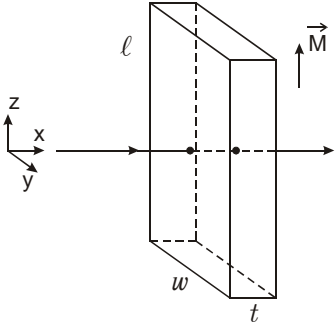}
\caption{\label{sample} Schematic layout of a homogeneously magnetized rectangular
sample.}
\end{figure}

In this appendix we will calculate the magnetic induction $\vec{B}$ generated
by a uniformly magnetized sample with length $l$, width $w$, and thickness $t$
(Fig.~\ref{sample}).  Moreover, analytical expressions will be given for the line integrals of
$\vec{B}$ along the path of a neutron. The center of the sample is taken as the origin of
the reference frame.

Our starting point is the Biot-Savart law:
\begin{equation} \vec{B}(x,y,z) = \frac{\mu_0}{4\pi} \int_S
\frac{\vec{M} \times \vec{n} \times \vec{r}}{r^3} dS + \frac{\mu_0}{4\pi}
\int_\tau \frac{\nabla \times \vec{M} \times \vec{r}}{r^3} d\tau
\end{equation} where $\mu_0 = 4\pi \times 10^{-7}$ H/m, $\vec{M}$ is the
magnetization, $\vec{n}$ the unit vector perpendicular to the surface $S$ of
the sample, $\vec{r}$ the vector pointing from the surface $S$ to the point
$(x,y,z)$, and $\tau$ the volume of the sample. Since the sample has a
homogeneous magnetization, the second term vanishes and, with $\vec{m} =
\vec{M} / M$, we have
\begin{equation} \vec{B} = \frac{\mu_0 M}{4\pi} \int_S
\frac{\vec{m} \times \vec{n} \times \vec{r}}{r^3} dS.
\end{equation}

A straightforward but tedious calculation shows that \begin{eqnarray} B_x & =
& - \sum_{\epsilon_1, \epsilon_2, \epsilon_3 = \pm 1} \epsilon_1 \epsilon_2
\epsilon_3 \ln \left( \mathcal{S} - \left( 2y + \epsilon_2 w \right) \right)
\nonumber \\ B_y & = & - \sum_{\epsilon_1, \epsilon_2, \epsilon_3 = \pm 1}
\epsilon_1 \epsilon_2 \epsilon_3 \ln \left( \mathcal{S} - \left( 2x +
\epsilon_1 t \right) \right) \\ B_z & = & \sum_{\epsilon_1, \epsilon_2,
\epsilon_3 = \pm 1} \epsilon_1 \epsilon_2 \epsilon_3 \arctan \left( \frac{2z +
\epsilon_3 l}{\left( 2x + \epsilon_1 t \right) \left( 2y + \epsilon_2 w
\right)} \; \mathcal{S} \right) \nonumber \end{eqnarray} where
\begin{equation} \mathcal{S} = \sqrt{ \left( 2x + \epsilon_1 t \right)^2 +
\left( 2y + \epsilon_2 w \right)^2 + \left( 2z + \epsilon_3 l \right)^2}.
\end{equation}

The rotation of the polarization of a neutron beam depends on the line integral of the
magnetic field along the neutron path. From the Larmor equation $\frac{d}{dt} \vec{P}(t)
= \gamma \vec{P}(t) \times \vec{B}(t)$, or equivalently $\frac{d}{dx} \vec{P}(x) = \left(
\gamma / v \right) \vec{P}(x) \times \vec{B}(x)$ where $\gamma = 1.83 \times 10^8 $
s$^{-1} $T$^{-1}$ the gyromagnetic ratio of the neutron and $v$ its velocity, we get the
general solution
\begin{equation} \label{larmorsolution} \vec{P}(x,y,z) = \left\{
\exp \left( \frac{\gamma}{v} \int_{-\infty}^{x} \overline{\overline{B}} (x',y,z) dx'
\right) \right\} \vec{P}(-\infty,y,z) \end{equation} where we have defined the magnetic
field tensor $\overline{\overline{B}}$ as
\begin{equation} \overline{\overline{B}} (x,y,z) = \left(
\begin{array}{ccc}
  0    & B_z  & -B_y \\
  -B_z & 0    & B_x \\
  B_y  & -B_x & 0 \\
\end{array} \right) (x,y,z) \end{equation} Thus, in order to
calculate the rotation of the neutron beam polarization due to a homogeneously magnetized
sample, the following line integrals are required: \begin{eqnarray} && \hspace{-0.5cm} X
(y,z) = \int_{-\infty}^{\infty} B_x (x',y,z) dx' = 0 \nonumber
\\
&& \hspace{-0.5cm} Y (y,z) = \int_{-\infty}^{\infty} B_y (x',y,z) dx' =
\nonumber
\\
&& \hspace{0.5cm} - \frac{t \mu_0 M}{4\pi} \sum_{\epsilon_2,
\epsilon_3} \epsilon_2 \epsilon_3 \ln \left( \left(2y+\epsilon_2
w\right)^2 + \left(2z+\epsilon_3 l\right)^2 \right) \nonumber
\\
&& \hspace{-0.5cm} Z (y,z) = \int_{-\infty}^{\infty} B_z (x',y,z) dx' =
\nonumber
\\
&& \hspace{0.5cm} \frac{t \mu_0 M}{2\pi} \sum_{\epsilon_2, \epsilon_3}
\epsilon_2\epsilon_3 \arctan \left( \frac{2z+\epsilon_3 l}{2y+\epsilon_2 w}
\right) \nonumber
\\ \label{lineintegralsx} \end{eqnarray}

For completeness we also give the line integrals in the case the neutron beam
is along the $z$-direction:
\begin{eqnarray}
&& X' (x,y) = \int_{-\infty}^{\infty} B_x (x,y,z') dz' = 0 \nonumber
\\
&& Y' (x,y) = \int_{-\infty}^{\infty} B_y (x,y,z') dz' = 0 \nonumber
\\
&& Z' (x,y) = \int_{-\infty}^{\infty} B_z (x,y,z') dz' = \mu_0 M \; l
\end{eqnarray}

From Eq.~\ref{larmorsolution} and the above line integrals we get for the
final polarization $\vec{P}(\infty,y,z) = \overline{\overline{D}} (y,z) \times
\vec{P}(-\infty,y,z)$ with $\overline{\overline{D}} (y,z)$ equal to
\begin{equation} \label{DepMatrix} \small \frac{1}{\Sigma} \! \left(\!\!\!
\begin{array}{ccc}
  \Sigma \cos a\sqrt{\Sigma} & Z\sqrt{\Sigma}\sin a\sqrt{\Sigma}  & -Y\sqrt{\Sigma }\sin a\sqrt{\Sigma } \\
  -Z\sqrt{\Sigma }\sin a\sqrt{\Sigma } & Z^{2}\cos a\sqrt{\Sigma }+Y^{2}  & YZ\left( 1-\cos a\sqrt{\Sigma}\right) \\
  Y\sqrt{\Sigma }\sin a\sqrt{\Sigma } & YZ\left( 1-\cos a\sqrt{\Sigma }\right) & Y^{2}\cos a\sqrt{\Sigma}+Z^{2} \\
\end{array} \!\!\! \right)
\end{equation} where $\Sigma(y,z) = Y^2 (y,z) + Z^2 (y,z)$ and $a = \gamma/v$.

Now Eq.~\ref{DepMatrix} relates the initial polarization to the final polarization for a
beam passing through the sample at $(y,z)$. For a neutron beam with finite cross section
the matrix $\overline{\overline{D}}$ should be integrated over the beam cross section. If
the integration is symmetric relative to the origin, then we can make use of the fact
that \begin{equation} B_y (\epsilon_1 x, \epsilon_2 y, \epsilon_3 z) = \epsilon_2
\epsilon_3 \; B_y (x,y,z),
\end{equation} where $\epsilon_1, \epsilon_2, \epsilon_3 = \pm 1$. This means that
$B_y(x,y,z)$ and therefore $Y(y,z)$ are antisymmetric with respect to $y$ and $z$.
Therefore, from Eq.~\ref{DepMatrix}, we only have to integrate the matrix
\begin{equation} \label{DepMatrix2} \small \frac{1}{\Sigma} \left(\!\!\!
\begin{array}{ccc}
  \Sigma \cos a\sqrt{\Sigma} & Z\sqrt{\Sigma}\sin a\sqrt{\Sigma}  & 0  \\
  -Z\sqrt{\Sigma }\sin a\sqrt{\Sigma } & Z^{2}\cos a\sqrt{\Sigma }+Y^{2}  & 0 \\
  0 & 0 & Y^{2}\cos a\sqrt{\Sigma}+Z^{2} \\
\end{array} \!\!\! \right) \end{equation} over the cross section of the
neutron beam.

An infinitely narrow neutron beam passing exactly through the middle of the sample will
only have its polarization rotated by $B_z(x,0,0)$ since $B_x(x,0,0)$ and $B_y(x,0,0)$
vanish. As long as $Y^2(y,z)$ is small compared to $Z^2(y,z)$, which is valid if $(y,z)$
is sufficiently far from the edges, Eq.~\ref{DepMatrix2} is a pure rotation matrix.

It is now possible to calculate the magnetization of the sample from the measured
rotation angle (Eq.~\ref{xigammaphi}). If no stray fields were present, the rotation
angle would be given by $t \mu_0 M \gamma / v$. However, the stray fields reduce the
rotation angle to $\gamma / v Z(y,z)$ with $Z(y,z)$ given in Eq.~\ref{lineintegralsx}. We
can therefore define the geometrical factor $\eta$ as
\begin{equation} \phi = \eta \mu_0 M t \gamma / v
\end{equation} where $\eta(y,z)$ is given by \begin{equation} \eta(y,z)
= \frac{1}{2\pi} \sum_{\epsilon_2, \epsilon_3}
\epsilon_2\epsilon_3 \arctan \left( \frac{2z+\epsilon_3
l}{2y+\epsilon_2 w} \right).
\end{equation} Since $\eta(0,0)$ is a saddle point ($\eta(0,z)$
has a local maximum and $\eta(y,0)$ a local minimum) an average over the cross section of
the neutron beam, centered around the middle of the sample, will yield a result very
close to the value of $\eta(0,0)$, which is given by
\begin{equation} \eta = \frac{2}{\pi} \arctan \left( \frac{l}{w}
\right). \label{eta} \end{equation}


\begin{thebibliography}{22}
\expandafter\ifx\csname natexlab\endcsname\relax\def\natexlab#1{#1}\fi
\expandafter\ifx\csname bibnamefont\endcsname\relax
  \def\bibnamefont#1{#1}\fi
\expandafter\ifx\csname bibfnamefont\endcsname\relax
  \def\bibfnamefont#1{#1}\fi
\expandafter\ifx\csname citenamefont\endcsname\relax
  \def\citenamefont#1{#1}\fi
\expandafter\ifx\csname url\endcsname\relax
  \def\url#1{\texttt{#1}}\fi
\expandafter\ifx\csname urlprefix\endcsname\relax\def\urlprefix{URL }\fi
\providecommand{\bibinfo}[2]{#2} \providecommand{\eprint}[2][]{\url{#2}}

\bibitem[{\citenamefont{Saxena et~al.}(2000)\citenamefont{Saxena, Agarwal,
  Ahilan, Grosche, Haselwimmer, Steiner, Pugh, Walker, Julian, Monthoux
  et~al.}}]{UGe2}
\bibinfo{author}{\bibfnamefont{S.~S.} \bibnamefont{Saxena}},
  \bibinfo{author}{\bibfnamefont{P.}~\bibnamefont{Agarwal}},
  \bibinfo{author}{\bibfnamefont{K.}~\bibnamefont{Ahilan}},
  \bibinfo{author}{\bibfnamefont{F.~M.} \bibnamefont{Grosche}},
  \bibinfo{author}{\bibfnamefont{R.~K.~W.} \bibnamefont{Haselwimmer}},
  \bibinfo{author}{\bibfnamefont{M.~J.} \bibnamefont{Steiner}},
  \bibinfo{author}{\bibfnamefont{E.}~\bibnamefont{Pugh}},
  \bibinfo{author}{\bibfnamefont{I.~R.} \bibnamefont{Walker}},
  \bibinfo{author}{\bibfnamefont{S.~R.} \bibnamefont{Julian}},
  \bibinfo{author}{\bibfnamefont{P.}~\bibnamefont{Monthoux}},
  \bibnamefont{et~al.}, \bibinfo{journal}{Nature (London)}
  \textbf{\bibinfo{volume}{406}}, \bibinfo{pages}{587} (\bibinfo{year}{2000}).

\bibitem[{\citenamefont{Aoki et~al.}(2001)\citenamefont{Aoki, Huxley,
  Ressouche, Braithwaite, Flouquet, Brison, Lhotel, and Paulsen}}]{URhGe}
\bibinfo{author}{\bibfnamefont{D.}~\bibnamefont{Aoki}},
  \bibinfo{author}{\bibfnamefont{A.}~\bibnamefont{Huxley}},
  \bibinfo{author}{\bibfnamefont{E.}~\bibnamefont{Ressouche}},
  \bibinfo{author}{\bibfnamefont{D.}~\bibnamefont{Braithwaite}},
  \bibinfo{author}{\bibfnamefont{J.}~\bibnamefont{Flouquet}},
  \bibinfo{author}{\bibfnamefont{J.-P.} \bibnamefont{Brison}},
  \bibinfo{author}{\bibfnamefont{E.}~\bibnamefont{Lhotel}}, \bibnamefont{and}
  \bibinfo{author}{\bibfnamefont{C.}~\bibnamefont{Paulsen}},
  \bibinfo{journal}{Nature (London)} \textbf{\bibinfo{volume}{413}},
  \bibinfo{pages}{613} (\bibinfo{year}{2001}).

\bibitem[{\citenamefont{Moncton et~al.}(1978)\citenamefont{Moncton, Shirane,
  Thomlinson, Ishikawa, and Fischer}}]{DyMo6S8}
\bibinfo{author}{\bibfnamefont{D.~E.} \bibnamefont{Moncton}},
  \bibinfo{author}{\bibfnamefont{G.}~\bibnamefont{Shirane}},
  \bibinfo{author}{\bibfnamefont{W.}~\bibnamefont{Thomlinson}},
  \bibinfo{author}{\bibfnamefont{M.}~\bibnamefont{Ishikawa}}, \bibnamefont{and}
  \bibinfo{author}{\bibfnamefont{O.}~\bibnamefont{Fischer}},
  \bibinfo{journal}{Phys. Rev. Lett.} \textbf{\bibinfo{volume}{41}},
  \bibinfo{pages}{1133} (\bibinfo{year}{1978}).

\bibitem[{\citenamefont{Majkrzak et~al.}(1979)\citenamefont{Majkrzak, Shirane,
  Thomlinson, Ishikawa, Fischer, and Moncton}}]{GdMo6S8}
\bibinfo{author}{\bibfnamefont{C.~F.} \bibnamefont{Majkrzak}},
  \bibinfo{author}{\bibfnamefont{G.}~\bibnamefont{Shirane}},
  \bibinfo{author}{\bibfnamefont{W.}~\bibnamefont{Thomlinson}},
  \bibinfo{author}{\bibfnamefont{M.}~\bibnamefont{Ishikawa}},
  \bibinfo{author}{\bibfnamefont{O.}~\bibnamefont{Fischer}}, \bibnamefont{and}
  \bibinfo{author}{\bibfnamefont{D.~E.} \bibnamefont{Moncton}},
  \bibinfo{journal}{Solid State Commun.} \textbf{\bibinfo{volume}{31}},
  \bibinfo{pages}{773} (\bibinfo{year}{1979}).

\bibitem[{\citenamefont{Thomlinson et~al.}(1981)\citenamefont{Thomlinson,
  Shirane, Moncton, Ishikawa, and Fischer}}]{TbMo6S8}
\bibinfo{author}{\bibfnamefont{W.}~\bibnamefont{Thomlinson}},
  \bibinfo{author}{\bibfnamefont{G.}~\bibnamefont{Shirane}},
  \bibinfo{author}{\bibfnamefont{D.~E.} \bibnamefont{Moncton}},
  \bibinfo{author}{\bibfnamefont{M.}~\bibnamefont{Ishikawa}}, \bibnamefont{and}
  \bibinfo{author}{\bibfnamefont{O.}~\bibnamefont{Fischer}},
  \bibinfo{journal}{Phys. Rev. B} \textbf{\bibinfo{volume}{23}},
  \bibinfo{pages}{4455} (\bibinfo{year}{1981}).

\bibitem[{\citenamefont{Walker et~al.}(1997)\citenamefont{Walker, Grosche,
  Freye, and Lonzarich}}]{CeIn3}
\bibinfo{author}{\bibfnamefont{I.~R.} \bibnamefont{Walker}},
  \bibinfo{author}{\bibfnamefont{F.~M.} \bibnamefont{Grosche}},
  \bibinfo{author}{\bibfnamefont{D.~M.} \bibnamefont{Freye}}, \bibnamefont{and}
  \bibinfo{author}{\bibfnamefont{G.~G.} \bibnamefont{Lonzarich}},
  \bibinfo{journal}{Physica C} \textbf{\bibinfo{volume}{282}},
  \bibinfo{pages}{303} (\bibinfo{year}{1997}).

\bibitem[{\citenamefont{Grosche et~al.}(1997)\citenamefont{Grosche, Lister,
  Carter, Saxena, Haselwimmer, Mathur, Julian, and Lonzarich}}]{CePd2Si2}
\bibinfo{author}{\bibfnamefont{F.~M.} \bibnamefont{Grosche}},
  \bibinfo{author}{\bibfnamefont{S.~J.~S.} \bibnamefont{Lister}},
  \bibinfo{author}{\bibfnamefont{F.~V.} \bibnamefont{Carter}},
  \bibinfo{author}{\bibfnamefont{S.~S.} \bibnamefont{Saxena}},
  \bibinfo{author}{\bibfnamefont{R.~K.~W.} \bibnamefont{Haselwimmer}},
  \bibinfo{author}{\bibfnamefont{N.~D.} \bibnamefont{Mathur}},
  \bibinfo{author}{\bibfnamefont{S.~R.} \bibnamefont{Julian}},
  \bibnamefont{and} \bibinfo{author}{\bibfnamefont{G.~G.}
  \bibnamefont{Lonzarich}}, \bibinfo{journal}{Physica B}
  \textbf{\bibinfo{volume}{239}}, \bibinfo{pages}{62} (\bibinfo{year}{1997}).

\bibitem[{\citenamefont{Sato et~al.}(2001)\citenamefont{Sato, Aso, Miyake,
  Shiina, Thalmeier, Varelogiannis, Geibel, Steglich, Fulde, and
  Komatsubara}}]{UPd2Al3}
\bibinfo{author}{\bibfnamefont{N.~K.} \bibnamefont{Sato}},
  \bibinfo{author}{\bibfnamefont{N.}~\bibnamefont{Aso}},
  \bibinfo{author}{\bibfnamefont{K.}~\bibnamefont{Miyake}},
  \bibinfo{author}{\bibfnamefont{R.}~\bibnamefont{Shiina}},
  \bibinfo{author}{\bibfnamefont{P.}~\bibnamefont{Thalmeier}},
  \bibinfo{author}{\bibfnamefont{G.}~\bibnamefont{Varelogiannis}},
  \bibinfo{author}{\bibfnamefont{C.}~\bibnamefont{Geibel}},
  \bibinfo{author}{\bibfnamefont{F.}~\bibnamefont{Steglich}},
  \bibinfo{author}{\bibfnamefont{P.}~\bibnamefont{Fulde}}, \bibnamefont{and}
  \bibinfo{author}{\bibfnamefont{T.}~\bibnamefont{Komatsubara}},
  \bibinfo{journal}{Nature (London)} \textbf{\bibinfo{volume}{410}},
  \bibinfo{pages}{340} (\bibinfo{year}{2001}).

\bibitem[{\citenamefont{Fertig et~al.}(1977)\citenamefont{Fertig, Johnston,
  DeLong, McCallum, Maple, and Matthias}}]{ErRh4B4_1}
\bibinfo{author}{\bibfnamefont{W.~A.} \bibnamefont{Fertig}},
  \bibinfo{author}{\bibfnamefont{D.~C.} \bibnamefont{Johnston}},
  \bibinfo{author}{\bibfnamefont{L.~E.} \bibnamefont{DeLong}},
  \bibinfo{author}{\bibfnamefont{R.~W.} \bibnamefont{McCallum}},
  \bibinfo{author}{\bibfnamefont{M.~B.} \bibnamefont{Maple}}, \bibnamefont{and}
  \bibinfo{author}{\bibfnamefont{B.~T.} \bibnamefont{Matthias}},
  \bibinfo{journal}{Phys. Rev. Lett.} \textbf{\bibinfo{volume}{38}},
  \bibinfo{pages}{987} (\bibinfo{year}{1977}).

\bibitem[{\citenamefont{Moncton et~al.}(1980)\citenamefont{Moncton, McWhan,
  Schmidt, Shirane, Thomlinson, Maple, MacKay, Woolf, Fisk, and
  Johnston}}]{ErRh4B4_2}
\bibinfo{author}{\bibfnamefont{D.~E.} \bibnamefont{Moncton}},
  \bibinfo{author}{\bibfnamefont{D.~B.} \bibnamefont{McWhan}},
  \bibinfo{author}{\bibfnamefont{P.~H.} \bibnamefont{Schmidt}},
  \bibinfo{author}{\bibfnamefont{G.}~\bibnamefont{Shirane}},
  \bibinfo{author}{\bibfnamefont{W.}~\bibnamefont{Thomlinson}},
  \bibinfo{author}{\bibfnamefont{M.~B.} \bibnamefont{Maple}},
  \bibinfo{author}{\bibfnamefont{H.~B.} \bibnamefont{MacKay}},
  \bibinfo{author}{\bibfnamefont{L.~D.} \bibnamefont{Woolf}},
  \bibinfo{author}{\bibfnamefont{Z.}~\bibnamefont{Fisk}}, \bibnamefont{and}
  \bibinfo{author}{\bibfnamefont{D.~C.} \bibnamefont{Johnston}},
  \bibinfo{journal}{Phys. Rev. Lett.} \textbf{\bibinfo{volume}{45}},
  \bibinfo{pages}{2060} (\bibinfo{year}{1980}).

\bibitem[{\citenamefont{Ishikawa and Fischer}(1977)}]{HoMo6S8}
\bibinfo{author}{\bibfnamefont{M.}~\bibnamefont{Ishikawa}} \bibnamefont{and}
  \bibinfo{author}{\bibfnamefont{O.}~\bibnamefont{Fischer}},
  \bibinfo{journal}{Solid State Commun.} \textbf{\bibinfo{volume}{23}},
  \bibinfo{pages}{37} (\bibinfo{year}{1977}).

\bibitem[{\citenamefont{Tateiwa et~al.}(2001)\citenamefont{Tateiwa, Kobayashi,
  Hanazono, Amaya, Haga, Settai, and Onuki}}]{Tateiwa}
\bibinfo{author}{\bibfnamefont{N.}~\bibnamefont{Tateiwa}},
  \bibinfo{author}{\bibfnamefont{T.~C.} \bibnamefont{Kobayashi}},
  \bibinfo{author}{\bibfnamefont{K.}~\bibnamefont{Hanazono}},
  \bibinfo{author}{\bibfnamefont{K.}~\bibnamefont{Amaya}},
  \bibinfo{author}{\bibfnamefont{Y.}~\bibnamefont{Haga}},
  \bibinfo{author}{\bibfnamefont{R.}~\bibnamefont{Settai}}, \bibnamefont{and}
  \bibinfo{author}{\bibfnamefont{Y.}~\bibnamefont{Onuki}}, \bibinfo{journal}{J.
  Phys. Condens. Matter} \textbf{\bibinfo{volume}{13}}, \bibinfo{pages}{L17}
  (\bibinfo{year}{2001}).

\bibitem[{\citenamefont{Nishioka et~al.}(2002)\citenamefont{Nishioka, Motoyama,
  Nakamura, Kadoya, and Sato}}]{Nishioka}
\bibinfo{author}{\bibfnamefont{T.}~\bibnamefont{Nishioka}},
  \bibinfo{author}{\bibfnamefont{G.}~\bibnamefont{Motoyama}},
  \bibinfo{author}{\bibfnamefont{S.}~\bibnamefont{Nakamura}},
  \bibinfo{author}{\bibfnamefont{H.}~\bibnamefont{Kadoya}}, \bibnamefont{and}
  \bibinfo{author}{\bibfnamefont{N.~K.} \bibnamefont{Sato}},
  \bibinfo{journal}{Phys. Rev. Lett.} \textbf{\bibinfo{volume}{88}},
  \bibinfo{pages}{237203} (\bibinfo{year}{2002}).

\bibitem[{\citenamefont{Nishioka et~al.}(2003)\citenamefont{Nishioka, Motoyama,
  Nakamura, Kadoya, and Sato}}]{Nishioka2}
\bibinfo{author}{\bibfnamefont{T.}~\bibnamefont{Nishioka}},
  \bibinfo{author}{\bibfnamefont{N.~K.} \bibnamefont{Sato}}, \bibnamefont{and}
  \bibinfo{author}{\bibfnamefont{G.}~\bibnamefont{Motoyama}},
  \bibinfo{journal}{Phys. Rev. Lett.} \textbf{\bibinfo{volume}{91}},
  \bibinfo{pages}{209702} (\bibinfo{year}{2003}).

\bibitem[{\citenamefont{Boulet et~al.}(1997)\citenamefont{Boulet, Daoudi,
  Potel, Noel, Gross, Andre, and Bouree}}]{Boulet}
\bibinfo{author}{\bibfnamefont{P.}~\bibnamefont{Boulet}},
  \bibinfo{author}{\bibfnamefont{A.}~\bibnamefont{Daoudi}},
  \bibinfo{author}{\bibfnamefont{M.}~\bibnamefont{Potel}},
  \bibinfo{author}{\bibfnamefont{H.}~\bibnamefont{Noel}},
  \bibinfo{author}{\bibfnamefont{G.~M.} \bibnamefont{Gross}},
  \bibinfo{author}{\bibfnamefont{G.}~\bibnamefont{Andre}}, \bibnamefont{and}
  \bibinfo{author}{\bibfnamefont{F.}~\bibnamefont{Bouree}},
  \bibinfo{journal}{J. Alloys Comp.} \textbf{\bibinfo{volume}{247}},
  \bibinfo{pages}{104} (\bibinfo{year}{1997}).

\bibitem[{\citenamefont{Onuki et~al.}(1992)\citenamefont{Onuki, Ukon, Yun,
  Umehara, Satoh, Fukuhara, Sato, Takayanagi, Shikama, and Ochiai}}]{Onuki}
\bibinfo{author}{\bibfnamefont{Y.}~\bibnamefont{Onuki}},
  \bibinfo{author}{\bibfnamefont{I.}~\bibnamefont{Ukon}},
  \bibinfo{author}{\bibfnamefont{S.~W.} \bibnamefont{Yun}},
  \bibinfo{author}{\bibfnamefont{I.}~\bibnamefont{Umehara}},
  \bibinfo{author}{\bibfnamefont{K.}~\bibnamefont{Satoh}},
  \bibinfo{author}{\bibfnamefont{T.}~\bibnamefont{Fukuhara}},
  \bibinfo{author}{\bibfnamefont{H.}~\bibnamefont{Sato}},
  \bibinfo{author}{\bibfnamefont{S.}~\bibnamefont{Takayanagi}},
  \bibinfo{author}{\bibfnamefont{M.}~\bibnamefont{Shikama}}, \bibnamefont{and}
  \bibinfo{author}{\bibfnamefont{A.}~\bibnamefont{Ochiai}},
  \bibinfo{journal}{J. Phys. Soc. Jpn.} \textbf{\bibinfo{volume}{61}},
  \bibinfo{pages}{293} (\bibinfo{year}{1992}).

\bibitem[{\citenamefont{Huxley et~al.}(2000)\citenamefont{Huxley, Sheikin, and
  Braithwaite}}]{Huxley}
\bibinfo{author}{\bibfnamefont{A.}~\bibnamefont{Huxley}},
  \bibinfo{author}{\bibfnamefont{I.}~\bibnamefont{Sheikin}}, \bibnamefont{and}
  \bibinfo{author}{\bibfnamefont{D.}~\bibnamefont{Braithwaite}},
  \bibinfo{journal}{Physica B} \textbf{\bibinfo{volume}{284-288}},
  \bibinfo{pages}{1277} (\bibinfo{year}{2000}).

\bibitem[{\citenamefont{Rekveldt}(1973)}]{Rekveldt}
\bibinfo{author}{\bibfnamefont{M.~T.} \bibnamefont{Rekveldt}},
  \bibinfo{journal}{Z. Phys.} \textbf{\bibinfo{volume}{259}},
  \bibinfo{pages}{391} (\bibinfo{year}{1973}).

\bibitem[{\citenamefont{Rosman and Rekveldt}(1990)}]{Rosman1}
\bibinfo{author}{\bibfnamefont{R.}~\bibnamefont{Rosman}} \bibnamefont{and}
  \bibinfo{author}{\bibfnamefont{M.~T.} \bibnamefont{Rekveldt}},
  \bibinfo{journal}{Z. Phys. B} \textbf{\bibinfo{volume}{79}},
  \bibinfo{pages}{61} (\bibinfo{year}{1990}).

\bibitem[{\citenamefont{Rosman and Rekveldt}(1991)}]{Rosman2}
\bibinfo{author}{\bibfnamefont{R.}~\bibnamefont{Rosman}} \bibnamefont{and}
  \bibinfo{author}{\bibfnamefont{M.~T.} \bibnamefont{Rekveldt}},
  \bibinfo{journal}{Phys. Rev. B} \textbf{\bibinfo{volume}{43}},
  \bibinfo{pages}{8437} (\bibinfo{year}{1991}).

\bibitem[{\citenamefont{Pfleiderer and Huxley}(2002)}]{Pleiderer}
\bibinfo{author}{\bibfnamefont{C.}~\bibnamefont{Pfleiderer}} \bibnamefont{and}
  \bibinfo{author}{\bibfnamefont{A.~D.} \bibnamefont{Huxley}},
  \bibinfo{journal}{Phys. Rev. Lett.} \textbf{\bibinfo{volume}{89}},
  \bibinfo{pages}{147005} (\bibinfo{year}{2002}).

\bibitem[{\citenamefont{Lhotel et~al.}(2003)\citenamefont{Lhotel, Paulsen, and
  Huxley}}]{Lhotel}
\bibinfo{author}{\bibfnamefont{E.}~\bibnamefont{Lhotel}},
  \bibinfo{author}{\bibfnamefont{C.}~\bibnamefont{Paulsen}}, \bibnamefont{and}
  \bibinfo{author}{\bibfnamefont{A.~D.} \bibnamefont{Huxley}},
  \bibinfo{journal}{Phys. Rev. Lett.} \textbf{\bibinfo{volume}{91}},
  \bibinfo{pages}{209701} (\bibinfo{year}{2003}).

\end{thebibliography}
\end{document}